\documentstyle[twoside]{article}

\catcode`\@=11
\long\def\@makefntext#1{
\protect\noindent \hbox to 3.2pt {\hskip-.9pt  
$^{{\eightrm\@thefnmark}}$\hfil}#1\hfill}		

\def\@makefnmark{\hbox to 0pt{$^{\@thefnmark}$\hss}}	
	
\def\ps@myheadings{\let\@mkboth\@gobbletwo
\def\@oddhead{\hbox{}
\rightmark\hfil\eightrm\thepage}   
\def\@oddfoot{}\def\@evenhead{\eightrm\thepage\hfil
\leftmark\hbox{}}\def\@evenfoot{}
\def\sectionmark##1{}\def\subsectionmark##1{}}

\oddsidemargin=\evensidemargin
\newcounter{sectionc}\newcounter{subsectionc}\newcounter{subsubsectionc}
\renewcommand{\section}[1] {\vspace{12pt}\addtocounter{sectionc}{1} 
\setcounter{subsectionc}{0}\setcounter{subsubsectionc}{0}\noindent 
	{\tenbf\thesectionc. #1}\par\vspace{5pt}}
\renewcommand{\subsection}[1] {\vspace{12pt}\addtocounter{subsectionc}{1} 
	\setcounter{subsubsectionc}{0}\noindent 
	{\bf\thesectionc.\thesubsectionc. {\kern1pt \bfit #1}}\par\vspace{5pt}}
\renewcommand{\subsubsection}[1] {\vspace{12pt}\addtocounter{subsubsectionc}{1}
	\noindent{\tenrm\thesectionc.\thesubsectionc.\thesubsubsectionc.
	{\kern1pt \tenit #1}}\par\vspace{5pt}}
\newcommand{\nonumsection}[1] {\vspace{12pt}\noindent{\tenbf #1}
	\par\vspace{5pt}}

\newcounter{appendixc}
\newcounter{subappendixc}[appendixc]
\newcounter{subsubappendixc}[subappendixc]
\renewcommand{\thesubappendixc}{\Alph{appendixc}.\arabic{subappendixc}}
\renewcommand{\thesubsubappendixc}
	{\Alph{appendixc}.\arabic{subappendixc}.\arabic{subsubappendixc}}

\renewcommand{\appendix}[1] {\vspace{12pt}
        \refstepcounter{appendixc}
        \setcounter{figure}{0}
        \setcounter{table}{0}
        \setcounter{lemma}{0}
        \setcounter{theorem}{0}
        \setcounter{corollary}{0}
        \setcounter{definition}{0}
        \setcounter{equation}{0}
        \renewcommand{\thefigure}{\Alph{appendixc}.\arabic{figure}}
        \renewcommand{\thetable}{\Alph{appendixc}.\arabic{table}}
        \renewcommand{\theappendixc}{\Alph{appendixc}}
        \renewcommand{\thelemma}{\Alph{appendixc}.\arabic{lemma}}
        \renewcommand{\thetheorem}{\Alph{appendixc}.\arabic{theorem}}
        \renewcommand{\thedefinition}{\Alph{appendixc}.\arabic{definition}}
        \renewcommand{\thecorollary}{\Alph{appendixc}.\arabic{corollary}}
        \renewcommand{\theequation}{\Alph{appendixc}.\arabic{equation}}
        \noindent{\tenbf Appendix \theappendixc #1}\par\vspace{5pt}}
\newcommand{\subappendix}[1] {\vspace{12pt}
        \refstepcounter{subappendixc}
        \noindent{\bf Appendix \thesubappendixc. {\kern1pt \bfit #1}}
	\par\vspace{5pt}}
\newcommand{\subsubappendix}[1] {\vspace{12pt}
        \refstepcounter{subsubappendixc}
        \noindent{\rm Appendix \thesubsubappendixc. {\kern1pt \tenit #1}}
	\par\vspace{5pt}}

\newcommand{\textlineskip}{\baselineskip=13pt}
\newcommand{\smalllineskip}{\baselineskip=10pt}

\def\eightcirc{
\begin{picture}(0,0)
\put(4.4,1.8){\circle{6.5}}
\end{picture}}
\def\eightcopyright{\eightcirc\kern2.7pt\hbox{\eightrm c}} 

\newcommand{\copyrightheading}[1]
	{\vspace*{-2.5cm}\smalllineskip{\flushleft
	{\footnotesize International Journal of Modern Physics A, #1}\\
	{\footnotesize $\eightcopyright$\, World Scientific Publishing
	 Company}\\
	 }}

\def\abstracts#1#2#3{{
	\centering{\begin{minipage}{4.5in}\baselineskip=10pt\footnotesize
	\parindent=0pt #1\par 
	\parindent=15pt #2\par
	\parindent=15pt #3
	\end{minipage}}\par}} 

\renewenvironment{thebibliography}[1]
	{\frenchspacing
	 \ninerm\baselineskip=11pt
	 \begin{list}{\arabic{enumi}.}
	{\usecounter{enumi}\setlength{\parsep}{0pt}
	 \setlength{\leftmargin 12.7pt}{\rightmargin 0pt} 
	 \setlength{\itemsep}{0pt} \settowidth
	{\labelwidth}{#1.}\sloppy}}{\end{list}}

\newcounter{itemlistc}
\newcounter{romanlistc}
\newcounter{alphlistc}
\newcounter{arabiclistc}

\newcommand{\fcaption}[1]{
        \refstepcounter{figure}
        \setbox\@tempboxa = \hbox{\footnotesize Fig.~\thefigure. #1}
        \ifdim \wd\@tempboxa > 5in
           {\begin{center}
        \parbox{5in}{\footnotesize\smalllineskip Fig.~\thefigure. #1}
            \end{center}}
        \else
             {\begin{center}
             {\footnotesize Fig.~\thefigure. #1}
              \end{center}}
        \fi}

\newcommand{\tcaption}[1]{
        \refstepcounter{table}
        \setbox\@tempboxa = \hbox{\footnotesize Table~\thetable. #1}
        \ifdim \wd\@tempboxa > 5in
           {\begin{center}
        \parbox{5in}{\footnotesize\smalllineskip Table~\thetable. #1}
            \end{center}}
        \else
             {\begin{center}
             {\footnotesize Table~\thetable. #1}
              \end{center}}
        \fi}

\def\@citex[#1]#2{\if@filesw\immediate\write\@auxout
	{\string\citation{#2}}\fi
\def\@citea{}\@cite{\@for\@citeb:=#2\do
	{\@citea\def\@citea{,}\@ifundefined
	{b@\@citeb}{{\bf ?}\@warning
	{Citation `\@citeb' on page \thepage \space undefined}}
	{\csname b@\@citeb\endcsname}}}{#1}}

\newif\if@cghi
\def\cite{\@cghitrue\@ifnextchar [{\@tempswatrue
	\@citex}{\@tempswafalse\@citex[]}}
\def\citelow{\@cghifalse\@ifnextchar [{\@tempswatrue
	\@citex}{\@tempswafalse\@citex[]}}
\def\@cite#1#2{{$\null^{#1}$\if@tempswa\typeout
	{IJCGA warning: optional citation argument 
	ignored: `#2'} \fi}}

\def\pmb#1{\setbox0=\hbox{#1}
	\kern-.025em\copy0\kern-\wd0
	\kern.05em\copy0\kern-\wd0
	\kern-.025em\raise.0433em\box0}

\def\fnt#1#2{\footnotetext{\kern-.3em
	{$^{\mbox{\scriptsize #1}}$}{#2}}}

\def\fpage#1{\begingroup
\voffset=.3in
\thispagestyle{empty}\begin{table}[b]\centerline{\footnotesize #1}
	\end{table}\endgroup}

\def\runninghead#1#2{\pagestyle{myheadings}
\markboth{{\protect\footnotesize\it{\quad #1}}\hfill}
{\hfill{\protect\footnotesize\it{#2\quad}}}}
\font\tenrm=cmr10
\font\tenit=cmti10 
\font\tenbf=cmbx10
\font\bfit=cmbxti10 at 10pt
\font\ninerm=cmr9

\font\eightrm=cmr8

\def\qed{\hbox{${\vcenter{\vbox{			
   \hrule height 0.4pt\hbox{\vrule width 0.4pt height 6pt
   \kern5pt\vrule width 0.4pt}\hrule height 0.4pt}}}$}}

\newcommand{\s}{\sigma}
\renewcommand{\t}{\tau}
\newcommand{\p}{product integral}

\newcommand{\beq}{\begin{equation}}
\newcommand{\eeq}{\end{equation}}

\begin{document}

\runninghead{Product Integrals and Wilson loops}{Product Integrals and Wilson loops}

\normalsize\textlineskip
\thispagestyle{empty}
\setcounter{page}{1}

\copyrightheading{}                     

\vspace*{0.88truein}

\fpage{1}
\centerline{\bf PRODUCT INTEGRALS AND WILSON LOOPS}
\vspace*{0.37truein}
\centerline{\footnotesize ROBERT L. KARP
\footnote{Present
address: Department of Mathematics, Duke University, Durham, NC 27708.}}
\centerline{\footnotesize\it Physics Department, University
of Cincinnati}
\baselineskip=10pt
\centerline{\footnotesize\it Cincinnati, Ohio 45221-0011,
USA}
\vspace*{0.225truein}

\vspace*{0.21truein} \abstracts{ Using product integrals we review the
unambiguous mathematical representation of Wilson line and Wilson loop
operators, including their behavior under gauge transformations and the
non-abelian Stokes theorem. Interesting consistency conditions among
Wilson lines are also presented.} {}{}

\textlineskip                   
\vspace*{12pt}                  
\noindent

Wilson loops provide a convenient method of obtaining gauge invariant
observables from lattices to strings. On the other hand the product
integral formalism was developed in connection with matrix valued
differential equations\cite{rfour}, having built in the feature of {\it
order} of the matrices. Consequently these come quite handy when dealing
with {\it path ordered} quantities. As the theory of product integrals is
well founded, we can use the various results in the subject to study
Wilson lines and loops. In this paper we review the mathematics of the
formalism of product integrals and its applications to the physics of
Wilson lines and Wilson loops.

\section{Review of product integration}    
\vspace*{-0.5pt}
\noindent

The long history of {\em product integration} starts with
Volterra\cite{rfour}. In fact the \p\ is to the product what the ordinary
integral is to the sum. For starting point consider an evolution equation
of the type ${\bf Y'(s)} = A(s){\bf Y(x)}$ on the interval $[a,b]$. Given
${\bf Y}(a)$ we want to find ${\bf Y}(b)$. We can obtain an approximate
value
\beq {\bf Y}(b)\approx
e^{A(s_n)(s_n-s_{n-1})}\ldots e^{A(s_1) (s_1-s_0)}{\bf Y}(a)\equiv \Pi_P(A)
{\bf Y}(a) \label{e2} 
\eeq 
for any partition $P=\{s_0,s_1,\ldots,s_n\}$ of the interval $[a,b]$. In
fact this approximation process can be made completely precise, and in the
limit where the norm of the partition $\mu(P)$ goes to zero, under
suitable conditions, the expression (\ref{e2}) converges to the {\em \p }:
${\bf Y}(b)=\lim_{\mu(P)\rightarrow 0} \Pi_P(A) {\bf Y}(a)$.

We briefly enumerate some of the basic properties of product
integrals\cite{rfour,1}. First we note that $\prod_{a}^{x} e^{ A(s)ds}$ is
the solution of the initial value problem we started with. From the
formula $\det\left(\prod_{a}^{x} e^{ A(s)ds}\right)=e^{ \int_{a}^{x}\,
{\rm Tr} A(s)ds}$ we infer that product integrals are non-singular
matrices. We have a composition rule similar to that of ordinary
integrals: $\prod_{a}^{b} e^{ A(s)ds}=\prod_{c}^{b} e^{
A(s)ds}\prod_{a}^{c} e^{A(s)ds}$. The rule of differentiation with respect
to the endpoints reads:  $ \frac{\partial}{\partial x}\left(\prod_{y}^{x}
e^{ A(s)ds}\right) =A(x)\prod_{y}^{x} e^{ A(s)ds}$ and
$\frac{\partial}{\partial y}\left(\prod_{y}^{x} e^{ A(s)ds}\right) =
-\prod_{y}^{x} e^{ A(s)ds} A(y)$.

The usual way of computing ordinary integrals is by the means of the
Newton-Leibnitz formula, involving primitive functions.  The analogous
notion for product integrals uses the so-called L-operation. The
L-derivative for a nonsingular matrix valued function $P:[a,b]\rightarrow
{\bf C}_{n\times n}$ is defined by: $ LP(x):=P'(x)P^{-1}(x). $ It is
immediate that $\prod_{a}^{b} e^{ (LP)(s)ds}=P(b)P^{-1}(a)$, the analog of
the well known Newton-Leibnitz formula.

We reproduce three more important theorems that will be of great use later
on:

\noindent {\em Sum rule}: $ \prod_{a}^{x} e^{\,
[A(s)+B(s)]ds}=P(x)\prod_{a}^{x} e^{ \,P^{-1}(s)B(s)P(s)ds}$, where
$P(x)=\prod_{a}^{x} e^{ A(s)ds}$.

\noindent {\em Similarity rule}: $ P(x)\left( \prod_{a}^{x} e^{
B(s)ds}\right)P^{-1}(a)=\prod_{a}^{x} e^{\, [LP(s)+P(s)B(s)P^{-1}(s)]ds}.
$

\noindent {\em Derivative with respect to a parameter}:  For
$A(s;\lambda)$ with the right properties\cite{rfour} we have: $
\frac{\partial}{\partial \lambda}\prod_{y}^{x} e^{ A(s;\lambda)ds}
=\int_{y}^{x} d s\,\prod_{s}^{x} e^{ A(s;\lambda)ds}\frac{\partial
A}{\partial \lambda}(s;\lambda)\prod_{y}^{s} e^{ A(s;\lambda)ds}$.

\section{Wilson lines and Wilson loops}

Let $M$ be an n-dimensional manifold representing the space-time and $A$ a
Lie-algebra valued connection one-form $ A(x) = A_{\mu}^k (x)\; T_k \;
dx^{\mu}. $ For a curve $C$ (closed) the Wilson line (loop) is defined as
$ W[C] = {\cal P} e^{\int_{a}^{b} A}. $ If $C$ is parametrized by $\s$
then $A(x) = A(\s)d\s , $ where $ A(\s) \equiv
A^{\mu}(x(\s))dx^{\mu}/d\s$. $A(\s)$ plays the role of the matrix function
in the product integral.

Let us next consider the Wilson loop. For simplicity we assume that the
loop may be taken to be the boundary of a two dimensional orientable
surface $\Sigma$ parametrized by $(\s,\t)$. We can construct the
pulled-back field strength $F_{ab}$ in two different ways: as the the
field strength of the pulled-back connection or as the pull-back of the
field strength. It is easy to check that the two coincide.

Our goal is to express the Wilson loop operator in terms of product
integrals. As the Wilson loop depends only on the homotopy class of $C$,
we can parameterize $C$ conveniently: break it up into segments along which
either $\s$ or $\t$ remains constant. The composition rule for product
integrals ensures that the final result is independent of the intermediate
points. Accordingly we have $ W[C] = W_4\, W_3\, W_2\, W_1$. For these
intermediate Wilson lines $\s_0 = const.$ along $W_1$ and $W_3$, and $\s_1
= const.$ along $W_2$ and $W_4$.

Based on the properties of \p s, it was shown by two independent
methods\cite{1} that $W[C]$ has a surface integral representation of the
form:
\beq 
W=\prod_{\t_0}^{\t} e^{\int_{\s_0}^{\s} T^{-1}(\s';\t')
F_{01} (\s';\t') T(\s';\t')d\s' d\t'},
\label{st} 
\eeq 
where $T$ is a given Wilson line. This can be interpreted as a non-abelian
Stokes theorem.

Let us now focus on the transformation properties of Wilson lines and
Wilson loops under a gauge transformation $A (x) \longrightarrow g(x)A
(x)g^{-1}(x)-g(x)dg(x)^{-1}$. It was shown\cite{1} using product
integration that Wilson lines have the expected transformation ${\cal
P}e^{\int_a^b A}\longrightarrow g(b)\left( {\cal P}e^{\int_a^b
A}\right)g(a)$. For the Wilson loop one has two ways to derive the
transformation property: use the already established one for the lines, or
use the surface integral representation provided by the non-abelian Stokes
theorem (\ref{st}). It was checked\cite{1} that both ways we get the well
established answer, reinforcing our confidence in the techniques used.

\section{Further properties}

It is interesting to use the \p \ formalism to derive interesting
identities among Wilson lines. For brevity we introduce the following
notation: $P(A;b,a)= \prod_a^b e^{A(s)ds}$. Observing that $\prod_a^b
e^{A(s)ds}= \prod_a^b e^{[\lambda A(s)+(1-\lambda)A(s)]ds}$ for any real
$\lambda$, based on the sum rule we can deduce immediately that
\beq
P(\lambda A;a,b)\,P(A;b,a)=\prod_a^b e^{P(\lambda A;a,s)(1-\lambda)A(s)
P(\lambda A;s,a)ds}.
\label{ss}
\eeq 
An interesting immediate application of (\ref{ss}) is the addition law we
obtain for $\lambda={1\over 2}$: $P(2A;b,a)=P(A;b,a)\,\prod_a^b
e^{P(A;a,s)A(s) P(A;s,a)ds}$.

Another application of (\ref{ss}) is by differentiating with respect to
$\lambda$. After some algebra, involving the formula for differentiation
with respect to a parameter recalled earlier, we obtain the following
result:
\begin{eqnarray}\nonumber
&\int_a^b\! ds\, \{P(\lambda A;a,s)A(s)P(\lambda A;s,b)P(A;b,a)-
P(\lambda A;s,b)
P(A;b,s)\\
\nonumber
&[-P(\lambda A;a,s)A(s)+(1-\lambda)\int_a^s\! dt\, [P(\lambda A;a,s)A(s)
P(\lambda A;s,t)A(t)P(\lambda A;t,s)\\
&+P(\lambda A;a,t)A(t)P(\lambda A;t,s)
A(s)]P(A;s,a)
]\}=0. 
\end{eqnarray}
In conclusion we note that 
the results described here were
extended to the supersymmetric case, marking the beginning of supersymmetric product 
integration\cite{3,4,5}.

\nonumsection{Acknowledgements}
\noindent 
Most of the results presented here were obtained in an
enjoyable collaboration with F. Mansouri. This work was supported in
part by the Department of Energy under the contract number
DOE-FGO2-84ER40153.

\nonumsection{References}

\end{document}